\documentclass[epj]{svjour}
\usepackage{graphicx}
\usepackage{bm}
\usepackage{amsmath}
\usepackage[usenames,dvipsnames,svgnames,table]{xcolor}

\usepackage{graphics}
\begin{document}
\title{Surface light modulation by sea ice and phytoplankton survival in a convective flow model}
\author{Vinicius Beltram Tergolina \and Enrico Calzavarini \and Gilmar Mompean \and Stefano Berti
}                     
%
%
\institute{Univ. Lille, ULR 7512 - Unit\'e de M\'ecanique de Lille Joseph Boussinesq (UML), F-59000 Lille, France}
\date{Received: date / Revised version: date}
%
\abstract{
Plankton dynamics depend in a complex manner on a variety of physical phenomena, according to both 
experimental and numerical data. In particular, experimental field studies have highlighted the relation 
between phytoplankton survival and turbulent upwelling and downwelling from thermal convection.
Recent numerical works have also shown the importance of accounting for advective transport by persistent 
structures in simulation models. In nutrient-rich polar marine environments phytoplankton blooms are critically 
limited by light availability under ice-covered waters. Such heterogeneity of the light intensity distribution, 
in association with a large-scale coherent fluid flow, can give rise  to nontrivial growth dynamics.
In this work we extend a previous advection-reaction-diffusion model of phytoplankton light-limited
vertical dynamics in the presence of convective transport. Specifically, we consider horizontally heterogeneous 
light conditions through the use of two regions with different production regimes, modelling the absence (presence) 
of light under (in between) obstacles. Such a model is intended as an idealized representation of nonuniformly 
ice-covered polar waters.
By means of numerical simulations, we find that the main role of advective transport is to hinder phytoplankton growth, 
but also that such effect depends on the positions of the obstacles with respect to the upwelling and downwelling flow regions.
Furthermore, we show that the sinking speed due to the density difference between phytoplankton organisms and water, 
while small, plays an important role, which depends on how it adds to the flow. 
These results indicate that advective transport can have a crucial impact on the survival conditions of
sinking phytoplankton species in polar environments.
}
\PACS{
      {PACS-key}{discribing text of that key}   \and
      {PACS-key}{discribing text of that key}
     } 

\maketitle

\section{Introduction}
\label{intro}

In phytoplankton ecology, biological and fluid dynamics are closely linked, ruling how planktonic populations 
interact with their environment and other organisms. Interdisciplinary work aimed at studying the interplay between 
plankton biology and the concurrent physical factors has allowed to considerably refine the  understanding of the
subject~\cite{ML2005,kiorboe2018mechanistic,prairie2012biophysical}. 
Notably, both modeling studies (see, e.g.,~\cite{taylor2011shutdown,lindemann2017dynamics,tergolina2021effects}) 
and direct measurements~\cite{lowry2018under} demonstrated the importance of vertical advective transport,
underlining the relation between turbulent convective fluid motions and phytoplankton growth dynamics.

Polar marine ecosystems are the seat of important plankton blooms (see~\cite{ardyna2020under} for a review). 
Understanding the dynamics of the latter is key 
to correctly estimate primary productivity~\cite{lowry2018under,arrigo2012massive,arrigo2015continued,assmy2017leads}, 
particularly under a changing climate scenario.
In the Arctic ocean, several studies reported under-ice phytoplankton blooms that were initiated by the onset 
of ice melt~\cite{lowry2018under,arrigo2012massive,assmy2017leads,boles2020under}.
Nevertheless, it is still not known whether under-ice blooms are controlled by the increase of light transmittance, 
leads (openings in the ice), upwelling from springtime convective mixing, biological factors, or other factors 
(most likely a combination of more than one).
Interestingly, however, it was observed that waters beneath sea ice with leads had significantly lower phytoplankton biomass, 
despite high nutrient availability, than regions without (or with few) leads~\cite{lowry2018under}.
It was argued that this phenomenon is controlled by the onset of convection below leads, which enhances mixing to larger depths, 
and reduces the travel time along the vertical. 
Consistently with Sverdrup's arguments~\cite{sverdrup1953conditions}, then, 
the bloom should be suppressed due to the mixed-layer depth being larger than the critical depth (where the integrated production 
is balanced by the integrated loss). 
Note that convection is here to be intended both as a thermo-solutal 
buoyancy-induced phenomenon, due to brine rejection from the ice and surface heat losses, 
and as wind-induced mixing in leads~\cite{lowry2018under}. 
It is worth remarking, too, 
that phytoplankton blooms beneath loosely consolidated ice
will likely become more common in the future Arctic ocean, as lead formation becomes more frequent due to increasingly thinner 
and dynamic ice coverage~\cite{lindsay2005thinning,serreze2007perspectives,kwok2013arctic}. 
This could significantly alter primary production, and have important consequences on local marine food webs.

In this work we take an approach based on idealized models, to explore the effect on phytoplankton dynamics of the horizontal modulation 
of light intensity at the surface of a vertical fluid layer. For this purpose, we extend a previous two-dimensional (2D) 
advection-reaction-diffusion model that we used to investigate the role of large and small scale fluid motions 
on the same biological system, but with uniform surface light intensity~\cite{tergolina2021effects}.
The role of vertical turbulent diffusivity on light-limited phytoplankton dynamics has been extensively studied 
in the past~\cite{shigesada1981analysis,huisman1994light,huisman2002sinking,huisman2002population} in one-dimensional (1D) systems. 
Our model can be seen as a 2D generalization of the one introduced in~\cite{huisman2002sinking}. Moreover, we also consider 
the presence of a coherent large-scale flow (as in~\cite{tergolina2021effects}) and of horizontally varying incident light 
intensity at the surface.
The flow field will be given in terms of a streamfunction corresponding to a kinematic cellular flow. 
While this is clearly a strong simplification with respect to a truly dynamic velocity field, it allows us 
to use realistic biological parameters and domain size. The cellular flow structure is intended as a simplified description 
of buoyancy and wind-driven flows~\cite{thorpe2005turbulent}, such as convective motions and Langmuir 
circulations~\cite{craik1976rational,bees1997planktonic,thorpe2004langmuir}, often encountered in the upper ocean. 
Beyond the environmental implications discussed above, from a more fundamental point of view, based on the overwhelming 
importance of the large-scale flow and of its downwelling region that we found in a previous work~\cite{tergolina2021effects}, 
in this study we wanted to see how phyotplankton dynamics change when reproduction is forbidden over either the upwelling or 
the downwelling flow region. As also in this case we expect a weak effect of small-scale flow features, 
we only consider a steady large-scale (cellular) flow. We then place a light-blocking obstacle, either over the downwelling region 
or over the upwelling one.

Through this approach we aim at assessing the importance of the different transport mechanisms on phytoplankton survival, 
in the presence of such vertically and horizontally heterogeneous growth dynamics.
We also examine different values of the sinking speed, due to phytoplankton being typically heavier than water. 
This corresponds to considering different algal species, and appears interesting on the basis of the experimental evidences 
about the varied species composition of phytoplankton in under-ice blooms~\cite{arrigo2012massive}.

This article is organized as follows. The model is introduced in Sec.~\ref{sec:model}. In Sec.~\ref{sec:results} we report
the results of our numerical study, and we discuss the impact of the different transport terms, by separately considering them 
in a systematic way. The main outcome and the conclusions are presented in Sec.~\ref{sec:concl}.

\section{Model}
\label{sec:model}

\begin{figure}[th]
\centering
\includegraphics[scale=0.65]{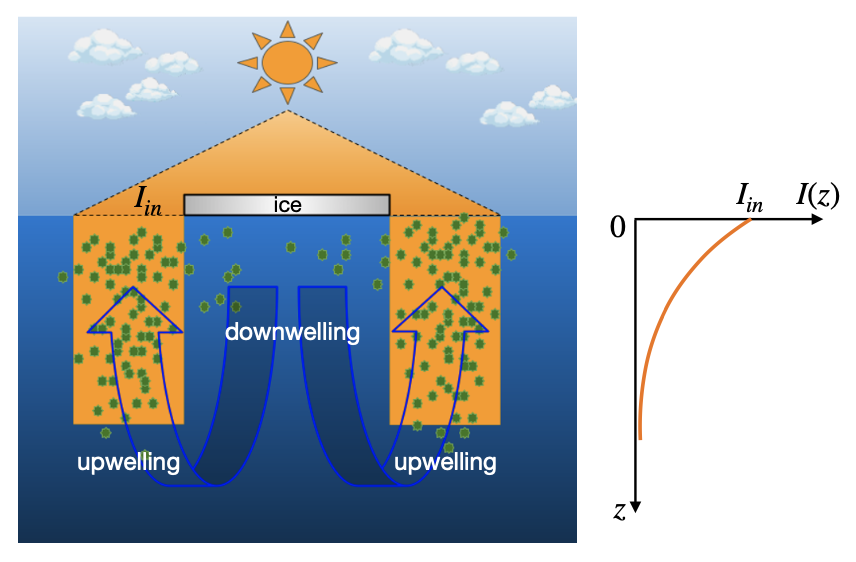}
\caption{Schematic of the model setup. The vertical decrease of light intensity from the surface value $I_{in}$ 
is described by Eq.~(\ref{eq:light_I}).Phytoplankton (green dots) dynamics are assumed to evolve according 
to Eqs.~(\ref{eq:fullunobs}) [with (\ref{eq:eq_prod}) and (\ref{eq:light_I})] and~(\ref{eq:fullobs}),
in the water columns in between the surface obstacles and below them (where $I_{in} = 0$), respectively. 
A cellular flow characterized by both upwelling and downwelling regions is also present.}
\label{fig:1}
\end{figure}

Our model is intended as an idealized representation of phytoplankton dynamics in waters that are partially covered
by ice. Specifically, we consider light-limited growth in a 2D vertical fluid layer, in the presence of obstacles 
at the surface mimicking ice blocks that do not allow the transmission of the incoming radiation to the water underneath. 
A schematic is provided in Fig.~\ref{fig:1}. While our description is an over-simplification of a realistic 
under-ice ecosystem, the present setup can provide indications useful to develop more complex biophysical models, 
and is in principle quite easily reproducible in laboratory experiments.

To describe the dynamics of the system is then necessary to specify the evolution equation of the population 
density field $\theta(x,z,t)$ (number of individuals per unit volume) in both the ice-free regions and in the ice-covered ones. 
In the regions receiving light (where phytoplankton can reproduce), we have:
\begin{equation}
    \frac{\partial \theta}{\partial t} = \left[ p(I) - l \right]\theta - \bm{v} \cdot \bm{\nabla} \theta + D \bm{\nabla}^{2}\theta, 
    \label{eq:fullunobs}
\end{equation}
as in~\cite{tergolina2021effects}, where $p(I)$ is the light-dependent production rate and $l$ the loss rate. 
In the regions below obstacles, instead, due to the absence of light, $p(I)=0$ and one has:
\begin{equation}
    \frac{\partial \theta}{\partial t} = - l\theta - \bm{v} \cdot \bm{\nabla} \theta + D \bm{\nabla}^{2}\theta .
    \label{eq:fullobs}
\end{equation}

The fluid layer in which we examine these dynamics is intended to represent the mixed layer (the water layer 
of nearly constant turbulent intensity found in upper oceans and lakes~\cite{vallis2017atmospheric}).
Its horizontal and vertical sizes will be denoted $L_x$ and $L_z$, respectively. The top and bottom boundaries 
are assumed to be rigid (as in~\cite{tergolina2021effects,huisman2002sinking}).

The production term in Eq.~(\ref{eq:fullunobs}) accounts for both water background turbidity, via the coefficient $\kappa_{bg}$, 
and population self-shading, with an attenuation factor $\kappa$. Its functional form is:
\begin{equation}
    p(I) = \frac{p_{max}I}{H+I},
    \label{eq:eq_prod}
\end{equation}
where $p_{max}$ is the maximum specific production rate, $H$ is a half-saturation constant and the time- and 
depth-dependent light-intensity is expressed as follows, according to Lambert-Beer's law:
\begin{equation}
    I(z,t) = I_{in}e^{-\int_{0}^{z}\kappa \theta(s , t)ds - \kappa_{bg}z},
    \label{eq:light_I}
\end{equation}
with $I_{in}$ the incident light (at the surface, where $z=0$). Note that we will assume $I_{in}=0$ in correspondence 
with the surface obstacles.
The biological parameter values used in our study, representative of realistic situations~\cite{huisman1999critical}, 
are reported in Table~\ref{tab:1}. They correspond to average measured values, with growth parameters obtained from 
freshwater phytoplankton species, and $\kappa_{bg}$ from clear lakes and coastal areas~\cite{huisman2002population}.
\begin{table*}[b!]
\centering
\caption{Parameters of the biological dynamics.}
\label{tab:1}
\begin{tabular}{ccc}
 \hline
 \hline
 \textbf{Parameter}& \textbf{Value}& \textbf{Meaning}\\ \hline
 $\kappa_{bg}$   & $0.2$~m$^{-1}$ & Background turbidity\\
 $\kappa$ &   $1.5 \cdot 10^{-11}$~m$^2$~cell$^{-1}$  & Specific light attenuation of phytoplankton\\
 $p_{max}$ & $0.04$~h$^{-1}$ & Maximal specific production rate\\
 $l$ & $0.01$~h$^{-1}$ & Specific loss rate\\
 $H$    & $30\,\mu$mol~photons~m$^{-2}$~s$^{-1}$ &  Half-saturation constant of light-limited growth\\
 $I_{in}$&   $350\,\mu$mol~photons~m$^{-2}$~s$^{-1}$  &Incident light intensity\\
 $v_{sink}$& $0.04$~m~h$^{-1}$  &Phytoplankton sinking velocity\\
 \hline
 \hline
\end{tabular}
\end{table*}

The coefficient $D$ in Eqs.~(\ref{eq:fullunobs}) and~(\ref{eq:fullobs}) represents an effective diffusivity, 
due to both small-scale unresolved turbulent motions and possible non-directed swimming behavior.
Advection is realized by a 2D incompressible flow $\bm{u}=(u_x,u_z)$ and phytoplankton is assumed to sink 
with a speed $v_{sink}\hat{\bm{z}}$, where $\hat{\bm{z}}$ is the unitary vector pointing downward in the vertical direction; 
the total velocity appearing in the evolution equations is thus $\bm{v}=\bm{u}+v_{sink}\hat{\bm{z}}$. 
For the phytoplankton sinking speed we take the value in Table~\ref{tab:1} as the reference one, and we will later vary it.  

For the velocity field we choose a cellular flow, which was originally introduced as a kinematic model of thermal convection 
fluid motions~\cite{solomon1988chaotic}. Its expression is obtained as $\bm{u}=\left(-\partial_z \Psi, \partial_x \Psi\right)$ 
from the following streamfunction:
\begin{equation}
    \Psi(x,z) = -\frac{U}{k}\sin(kx)\sin(kz),
    \label{eq:psi}
\end{equation}
where $U$ is the maximum velocity and $k = \pi/L_{z}$ a wavenumber. 
The flow pattern then corresponds to two counter-rotating steady vortices of size $L_z$, in the $(x,z)$ plane. 
It is worth stressing that this flow comprises
both a (central) downwelling region and two (side) upwelling ones, 
which typically play a relevant role on phytoplankton dynamics~\cite{lindemann2017dynamics,tergolina2021effects}.
Note that eddying motions on scales smaller than $L_z$, and unsteadiness, could be in principle taken into account by adding 
to the streamfunction of Eq.~(\ref{eq:psi}) analogous terms with larger wavenumbers, and an explicit time dependency 
(see, e.g.,~\cite{tergolina2021effects,lacorata2017chaotic,forgia2022numerical}). 
Our choice to neglect them, here, is motivated by a previous study 
without the surface light modulation~\cite{tergolina2021effects}, where we found that small unsteady eddies play a secondary role. 
Indeed, in that case, the fate (survival or extinction) of phytoplankton was primarily determined by the large-scale 
coherent component of the flow.

In the following we will consider two different geometrical configurations, corresponding to light-blocking obstacles, 
of fixed total size, positioned either above the downwelling flow region (LBD case), as in Fig.~\ref{fig:1}, 
or above the upwelling one (LBU case).
The horizontal and vertical domain sizes will be respectively set to $L_{x} = 60$~m and $L_{z} = 30$~m. 
Considering the chosen biological parameters (Table~\ref{tab:1}), 
a mixed layer depth of $30$~m is smaller 
than the critical depth (about $50$~m). Hence, in the absence of surface light heterogeneity and of a fluid flow, 
with the values of diffusivity explored, $1$~cm$^2$~s$^{-1} \leq D \leq 20$~cm$^2$~s$^{-1}$, which appear quite realistic 
for the ocean~\cite{tergolina2021effects,huisman2002sinking,huisman1999critical}, a bloom occurs. 
The presence of obstacles considerably reduces production, due to the absence of light below them. 
However, we made sure that also in this case phytoplankton blooms when $U=0$. 
This allowed us to investigate the impact of advection on the dynamics of the system, including a possible transition 
from population survival to extinction. 
Remark, too, that working with a relatively small system, but still apt to capture the salient dynamics, 
is computationally convenient.

We numerically study the dynamics of Eqs.~(\ref{eq:fullunobs}) and~(\ref{eq:fullobs}), which are simultaneously integrated 
in our rectangular domain (with $L_x=2L_z$) by means of a pseudo-Lagrangian algorithm~\cite{abel2001front,berti2005mixing} 
(see \cite{tergolina2021effects} for more details). 
We adopt periodic and no-flux boundary conditions along the horizontal ($x$) and the vertical ($z$), 
respectively (as in~\cite{tergolina2021effects}).
The initial condition is a low uniform population density  [$\theta(t=0) = 5.5 \cdot 10^{6}$~cells~m$^{-3}$]. 
To analyze the blooming conditions of the phytoplankton population we rely on the temporal behavior of the average biomass density, 
\begin{equation}
    \langle \theta \rangle(t) = \frac{1}{L_{x}L_{z}} \int_{0}^{L_{x}}\int_{0}^{L_{z}}\theta(x,z,t) \, dx dz,
    \label{eq:biomass}
\end{equation}
and the per-capita growth rate,
\begin{equation}
    r_p(t)=\frac{1}{\langle \theta \rangle}\frac{\partial \langle \theta \rangle}{\partial t}.
\label{eq:pc_gr}
\end{equation}
The per-capita growth rate is expected to attain a statistically constant value $r_p$ (after an initial transient), 
corresponding to exponential growth ($r_p>0$) or decay ($r_p<0$) in the early regime before the onset of nonlinear dynamical effects. 

\section{Results}
\label{sec:results} 

\begin{figure*}[t!]
\centering
\includegraphics[scale=0.42]{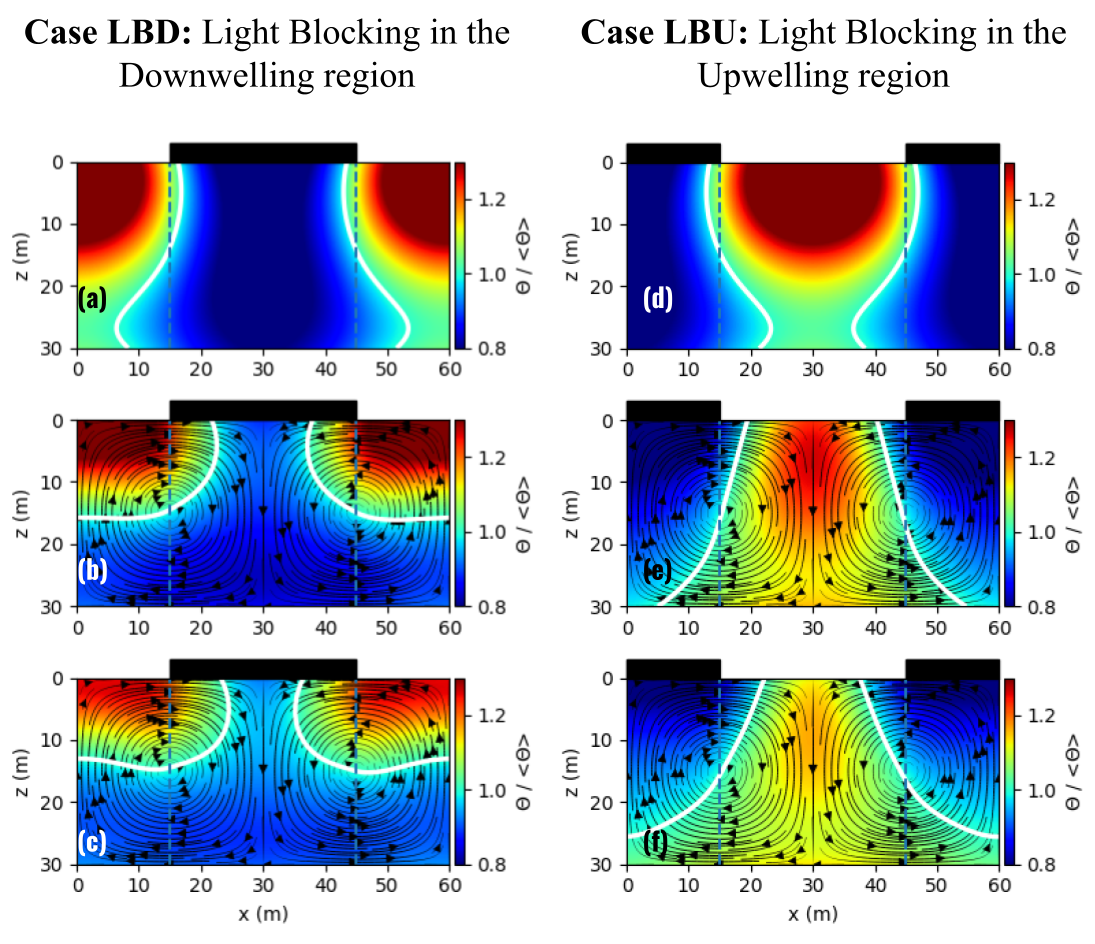}
\caption{Instantaneous normalized population density field $\theta(x,z,t^{*})/\langle \theta \rangle$ 
at a fixed instant of time $t^{*}=600$~h, for $D=10$~cm$^2$~s$^{-1}$ and $U=(0,0.71,1.19)$~m~h$^{-1}$ 
[panels (a,d), (b,e) and (c,f), respectively], where $\langle \theta \rangle$ stands for the spatial average. 
The obstacle is represented by the black rectangle, the vertical dashed lines delimit the light obstruction zone. 
The white lines are the isolines $\theta/\langle \theta \rangle = 1$, and $t^{*}$ is in the regime of stationary 
per-capita growth rate ($r_p(t)=r_p=\mathrm{const}$). The solid black lines represent flow streamlines, 
with arrows indicating the circulation direction.}
\label{fig:2}
\end{figure*}
We begin our numerical study by exploring the dynamics in two partially illuminated setups, both in the absence 
and in the presence of the fluid flow. 
As explained in Sec.~\ref{sec:model}, the blocking of light at the surface is realized by introducing obstacles 
of zero light transmittance, with different positions according to the cases (LBD, LBU) considered. Given the large number 
of control parameters (for both biology and transport) in the model, we decided to work with a single light-transmittance value 
associated with the obstacle. 
We then chose to use fully absorbing obstacles, in order to increase the contrast between the zones where 
light availability permits reproduction and those where this does not occur. 
Such a simplifying choice is also intended to ease the interpretation of the effects of the different transport 
and production regimes. Indeed, as we shall see, already in the present case the dynamics are quite complex. 
The light transmittance of ice is nevertheless variable and is a subject of debate (see~\cite{nicolaus2012changes} 
for recent measurements of ice thinning in the Arctic ocean). 

\begin{figure}[ht]
\centering
\includegraphics[scale=0.25]{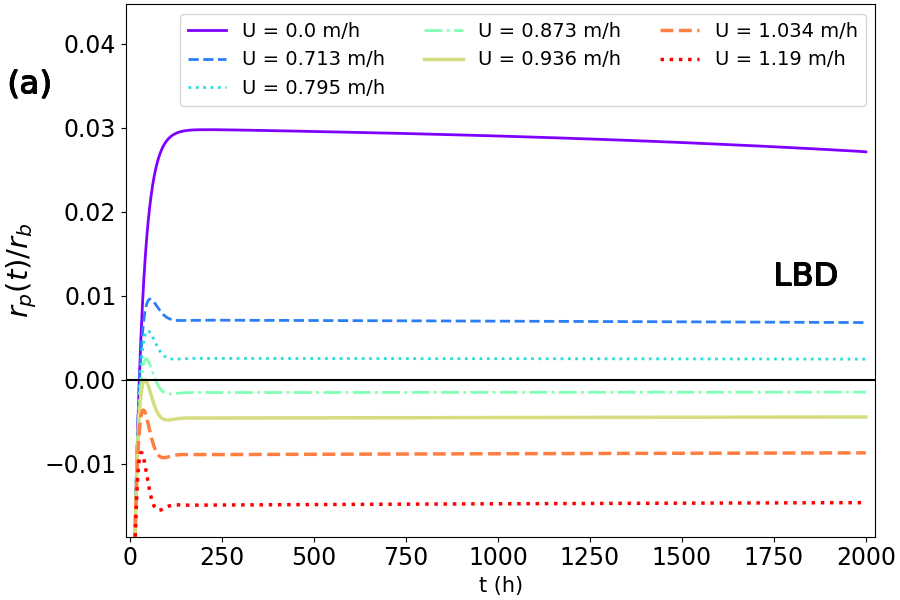}
\includegraphics[scale=0.25]{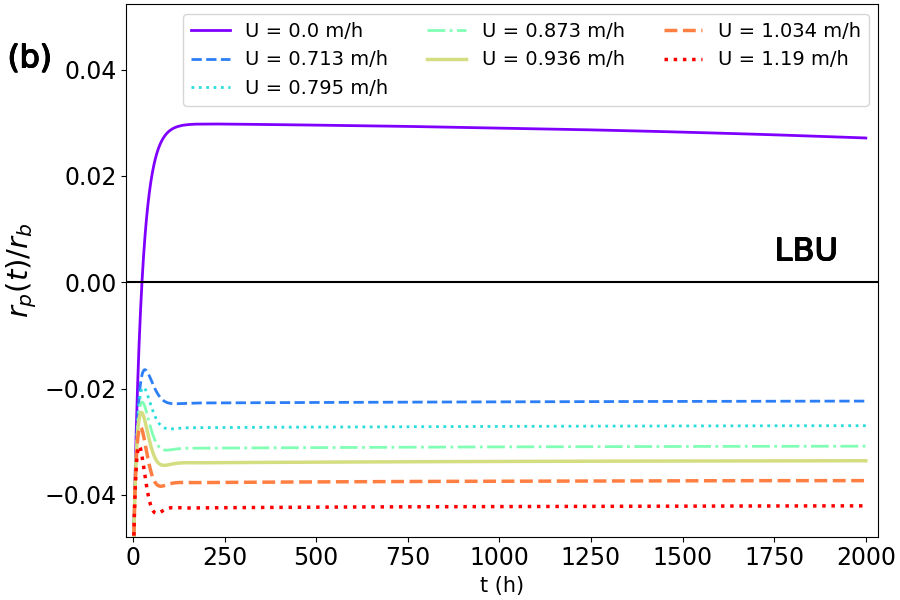}
\caption{Population per-capita growth rate $r_p(t)$, normalized by the intrinsic total (surface) growth rate $r_b$, 
as a function of time, for various values of $U$, $D=10$~cm$^2$~s$^{-1}$ and $L_z=30$~m. 
Panels (a) and (b) respectively refer to the LBD and LBU cases.}
\label{fig:3}
\end{figure}

Some snapshots of the field $\theta$ at a given time are shown in Fig.~\ref{fig:2}, for both the LBD and LBU cases,
and for increasing values of the flow intensity $U$.
This figure is obtained with the reference sinking speed $v_{sink}=0.04$~m~h$^{-1}$~\cite{tergolina2021effects,huisman2002sinking}. 
When advection is not present the two light-modulation cases give identical results, as expected. 
With nonzero advection, instead, important differences appear. 
While in the LBD case phytoplankton accumulates closer to the surface and presents small differences with respect to 
increasing flow intensity, in the LBU case the population spreads more in the vertical, reaching larger depths. 
By measuring the per-capita growth rate $r_p$ versus time for different values of $U$, we could assess that phytoplankton growth 
is hindered by advection, independently of the light-modulation considered. 
Indeed, the (constant) value of $r_p$ reached at large enough times, monotonically decreases with increasing $U$ 
in both the LBD and LBU cases [see Fig.~\ref{fig:3}, where $r_p$ is normalized by the total (birth minus death) 
growth rate at the surface $r_b=p_{max} I_{in}/(H+I_{in})-l$]. 
This effect parallels the one we already found in the absence of obstacles~\cite{tergolina2021effects}. 
However, the decrease of $r_p$ in the two situations is quantitatively different, and indeed larger when light does not 
penetrate in the upwelling region, so that an LBD/LBU asymmetry emerges. 
In the remainder of this work, we focus on the understanding of this feature. 

We can first observe that the orientation, with respect to the light gradient, of the fluid velocity alone
(which is opposite in the LBD and LBU cases) cannot be responsible for this effect. To see this, one can rewrite
Eq.~(\ref{eq:fullunobs}) as follows:
\begin{equation}
    \frac{\partial \theta}{\partial t} = g(\bm{x})\theta  - \bm{u} \cdot \bm{\nabla} \theta, 
    \label{eq:light_ar}
\end{equation}
where $g(\bm{x})$ indicates the spatially dependent growth rate, and diffusion and sinking have been neglected. 
Note that in our system the growth rate depends on both the vertical and the horizontal coordinates, due to the external 
light blocking and to self-shading. 
The fluid velocity $\bm{u}(\bm{x})$ depends on space, too, but not on time. 
In a Lagrangian reference frame the above equation reads:
\begin{eqnarray}
\begin{cases}
\dot{\theta} = g(\bm{x}(t))\theta,\\ 
\dot{\bm{x}}(t) = \bm{u}(\bm{x}(t)).
\end{cases}
\label{eq:light_arL}
\end{eqnarray}
Given an intial value for $\theta$ at time $t=0$ and a final position $\bm{x}(t)$, the solution of Eq.~(\ref{eq:light_arL}) is:
\begin{eqnarray}
\theta(\bm{x}(t),t)&=&\theta(\bm{x}(0),0)\ e^{\int_0^t g(\bm{x}(t')) dt'} \label{eq:arL_sol1}\\
&=&\theta(\bm{x}(0),0)\ e^{\int_{|\bm{x}(0)|}^{|\bm{x}(t)|} \frac{g(\bm{x}')}{|\bm{u}(\bm{x}')|} dx'}.
\label{eq:light_arL_sol}
\end{eqnarray}
It is therefore clear that the growth of $\theta$ along a given Lagrangian trajectory (coinciding with a streamline, 
due to flow stationarity) depends on the velocity at which the path is travelled but not on the orientation of the velocity itself. 
Moreover along a closed path, like a cellular-flow streamline, the variation of the reactive-scalar density 
will be independent of the circulation direction. A full loop starting with downward flow and followed by an upward flow 
will produce the same phytoplankton growth if it occurs in the reversed order.

The presence of a sinking speed, which modifies the Lagrangian equation of motion to
$\dot{\bm{x}}(t) = \bm{u}(\bm{x}(t)) + v_{sink} \hat{\bm{z}}$, can change the scenario because a descending flow path
will be covered faster than an ascending one, leading to a smaller growth for the first than for the latter.
Note that the presence of sinking also has an effect on the shapes of planktonic trajectories, in particular not all the trajectories
in the cellular flow will be closed~\cite{stommel1949trajectories}.
It is therefore the (slow) sinking movement that is responsible for breaking the symmetry of the growth dynamics between
the upwelling and downwelling flow regions.
The presence of diffusion will also act in a similar manner by allowing parcels of population density to cross the
flow streamlines. 

\begin{figure}[h!]
\centering
\includegraphics[scale=0.26]{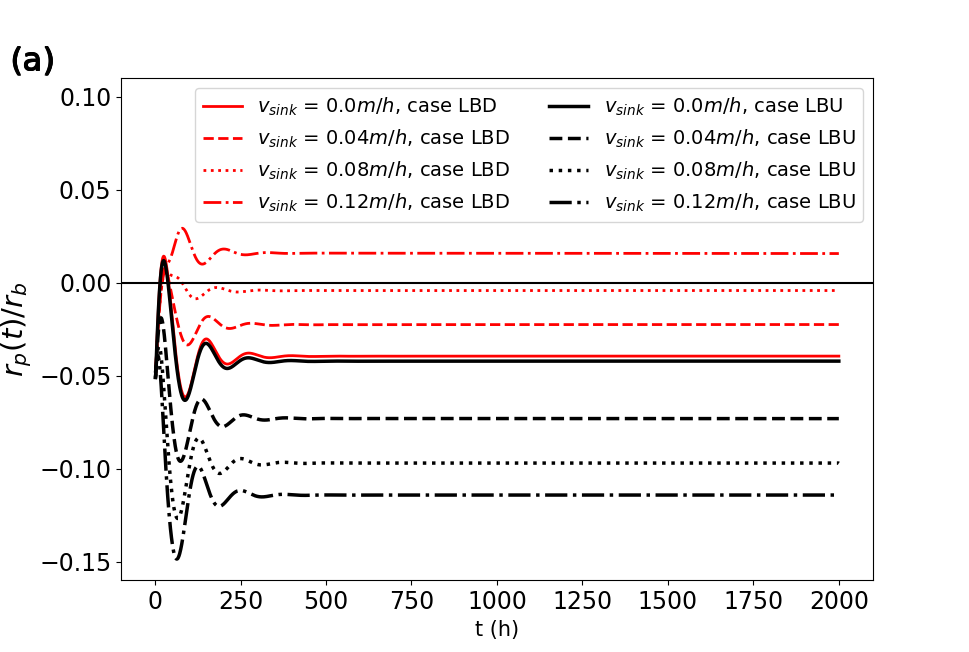}
\includegraphics[scale=0.26]{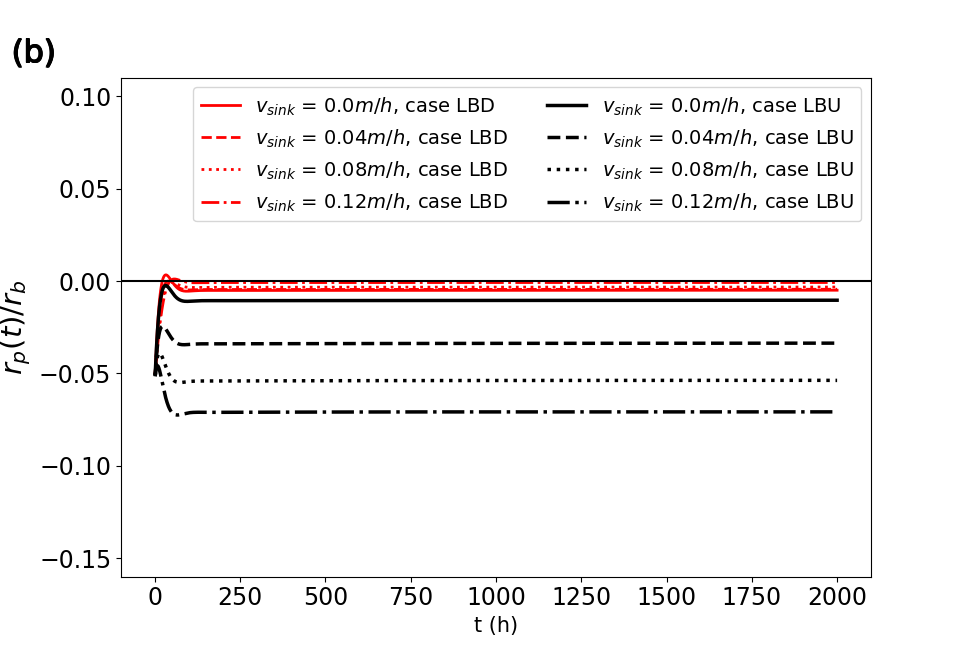}
\caption{Population per-capita growth rate, normalized by the intrinsic total (birth minus death) growth rate
at the surface, as a function of time for different values of the sinking speed in both the LBD case and the LBU one.
The flow intensity is $U = 0.936$~m/h, the diffusion coefficient is either $D = 1$~cm$^2$~s$^{-1}$ (a) or $D = 10$~cm$^2$~s$^{-1}$ (b).}
\label{fig:4}
\end{figure}

Figure~\ref{fig:4} reports the measure of the per-capita growth rate - normalized by the total (birth minus death) 
growth rate at the surface $r_b$ - versus time, for different values of the sinking speed and a fixed flow intensity. 
Panel (a) in the figure refers to a small value of the diffusion coefficient, while panel (b) is for a larger one. 
The role of diffusivity is quite subtle and we will come back to it later in this section. 
Independently of this, the results clearly show that in the absence of sinking ($v_{sink}=0$) the system is insensitive 
to the position of the obstacle. The value of $r_p$ is in fact the same, within numerical accuracy, in the LBD and LBU cases.
However, it decreases in the presence of sinking when light is blocked over the upwelling region, as it may be expected.
It is perhaps more surprising that it instead increases in the LBD case (which is more difficult to see for large $D$
in Fig.~\ref{fig:4}b due to the scale). For small diffusivity, it can even turn a no-bloom situation ($r_p<0$)
into a bloom one ($r_p>0$).

The combined effect of fluid advection, sinking and diffusion can be appreciated through the following argument, 
which helps rationalizing the previous observations. We consider the transformation (see also~\cite{ryabov2008population}):
\begin{equation}
\theta(\bm{x},t) = \textrm{exp}\left(
\frac{\bm{w}\cdot \bm{x}}{2 D}\right) \tilde{\theta}(\bm{x},t)
\label{eq:theta_transf_light}
\end{equation}
where $\bm{w}$ is a, yet to be defined, constant (in both time and space) velocity.
Plugging Eq.~(\ref{eq:theta_transf_light}) into Eq.~(\ref{eq:fullunobs}) we get:
\begin{equation}
    \frac{\partial \tilde{\theta}}{\partial t} = \left[ p(I) - l  - \frac{\bm{w}\cdot \bm{v}}{2 D} 
    + \frac{w^2}{4 D}\right]\tilde{\theta} - (\bm{v} -\bm{w}) \cdot \bm{\nabla} \tilde{\theta} + D \bm{\nabla}^{2}\tilde{\theta}. 
    \label{eq:fullunobs_trans1}
\end{equation}
Taking $\bm{w} = v_{sink} \hat{\bm{z}}$, and using the fact that $\bm{v} = \bm{u} + v_{sink} \hat{\bm{z}}$, we obtain
\begin{equation}
    \frac{\partial \tilde{\theta}}{\partial t} = \left[ p(I) - l  - \frac{v_{sink} u_z}{2 D} 
    - \frac{v_{sink}^2}{4 D}\right]\tilde{\theta} - \bm{u}  \cdot \bm{\nabla} \tilde{\theta} + D \bm{\nabla}^{2}\tilde{\theta}. 
    \label{eq:fullunobs_trans2}
\end{equation}
This is the evolution equation of a field advected by the fluid flow $\bm{u}$, where sinking is not explicitly present, 
but with a modified reaction term that now depends on the sign of the vertical velocity component $u_z$ (and also on space). 
It is thus clear that the local growth rate, i.e. $g = p(I) - l  - v_{sink} u_z/(2 D) - v_{sink}^2/(4 D)$, is always smaller 
in the downwelling regions ($u_z>0$) than in the upwelling ones ($u_z<0$). 
In short, $g_d < g_u$, where $d,u$ label down- and upwelling regions, respectively. 
We also note that the quadratic term in $v_{sink}$ can be neglected nearly everywhere, because in practice 
$v_{sink} \ll |u_z| \sim U$ (this may be not true close to the top and bottom boundaries, where $u_z$ also becomes very small). 
The above reasoning is independent of the presence of the obstacle blocking light at the surface. 
When the latter is present we can expect that the LBU situation is characterized by a growth rate similar to
the one observed in downwelling regions, $g_{LBU} \sim g_d$ and, on the opposite, for the LBD case: $g_{LBD} \sim g_u$.
This implies that $g_{LBD} > g_{LBU}$, in agreement with the observed trends.

One may then be tempted to extrapolate this result to the per-capita growth rate, 
$r_p=\partial_t \langle \theta \rangle / \langle \theta \rangle$, to obtain the following predictions 
for the two situations considered:
\begin{eqnarray}
\label{eq:rp_LBD}
r_{p,LBD} &\sim& \langle p(I) \rangle_L - l  + \frac{2 v_{sink} U}{\pi^2 D} 
    - \frac{v_{sink}^2}{4 D},\\
 \label{eq:rp_LBU}
r_{p,LBU} &\sim& \langle p(I) \rangle_L  - l  - \frac{2 v_{sink}  U}{\pi^2 D}
    - \frac{v_{sink}^2}{4 D},
\end{eqnarray}
which, however, assume a weak spatial coupling between the growth rate $g$ and $\tilde{\theta}$, and where we used 
the average vertical velocity in the regions receiving light, $\langle u_z \rangle_L  = \pm (2/\pi)^2 U$. 
More importantly, here we have neglected any other effects due to both advection and diffusion. 
In particular, this argument completely ignores the lateral transport, between the illuminated and shaded regions, 
due to both advection and diffusion.
The flux associated with this process is $u_x \theta - D \partial_x \theta$, with $u_x = \pm U cos(\pi z/ L_z)$ at the boundaries 
between the two regions, and it should be included to improve the present prediction. 
Indeed, the expressions in Eqs.~(\ref{eq:rp_LBD}) and (\ref{eq:rp_LBU}) fail to describe the behavior 
of the large-time value of $r_p$ as a function of $v_{sink}$, unless some fitting coefficients are introduced 
to account for the missing physics. While for small diffusivity the fitting coefficients are comparable in the LBD and LBU cases, 
which may suggest that the missing information resides in the $z$-dependence of $u_z$ and in the appropriate choice of the region 
where to average, this is not the case for large $D$. In fact, in this case, a striking feature that is already apparent 
from Fig.~\ref{fig:4}b is the asymmetric response of the system with respect to the position of the obstacle. 
This fact cannot be explained by Eqs.~(\ref{eq:rp_LBD}) and (\ref{eq:rp_LBU}), which are symmetric under 
the exchange of $U$ into $-U$.

Taking into account the role of lateral transport between shaded and illuminated regions analytically is not trivial.
Remark, too, that in the large $D$ limit analytical predictions are not available even in the 1D no-flow (fully illuminated)
model~\cite{huisman2002sinking}, starting from which ours was constructed.
In the following we then resort to an argument based on Lagrangian paths to provide a qualitative explanation of the asymmetric
response of the system.

\begin{figure}[htb]
\centering
\includegraphics[scale=0.55]{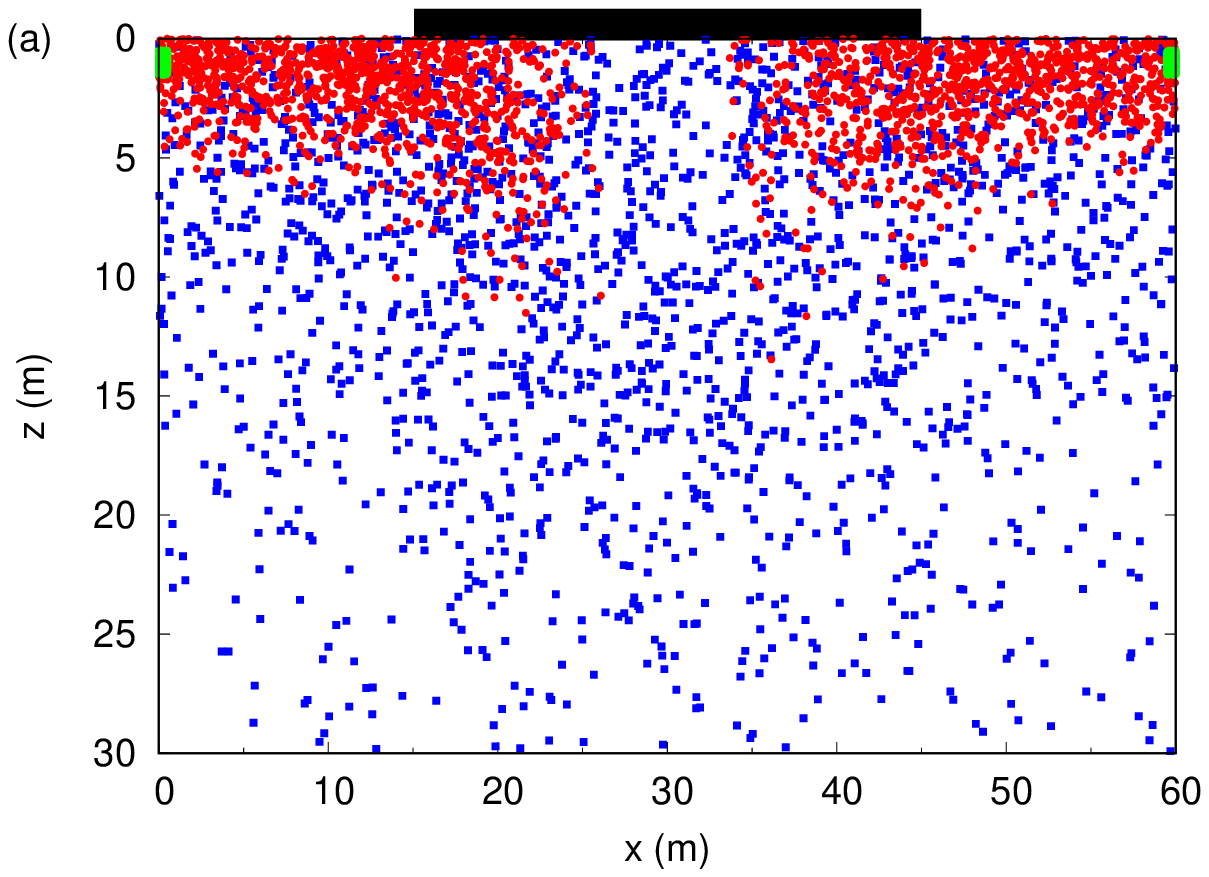}
\includegraphics[scale=0.55]{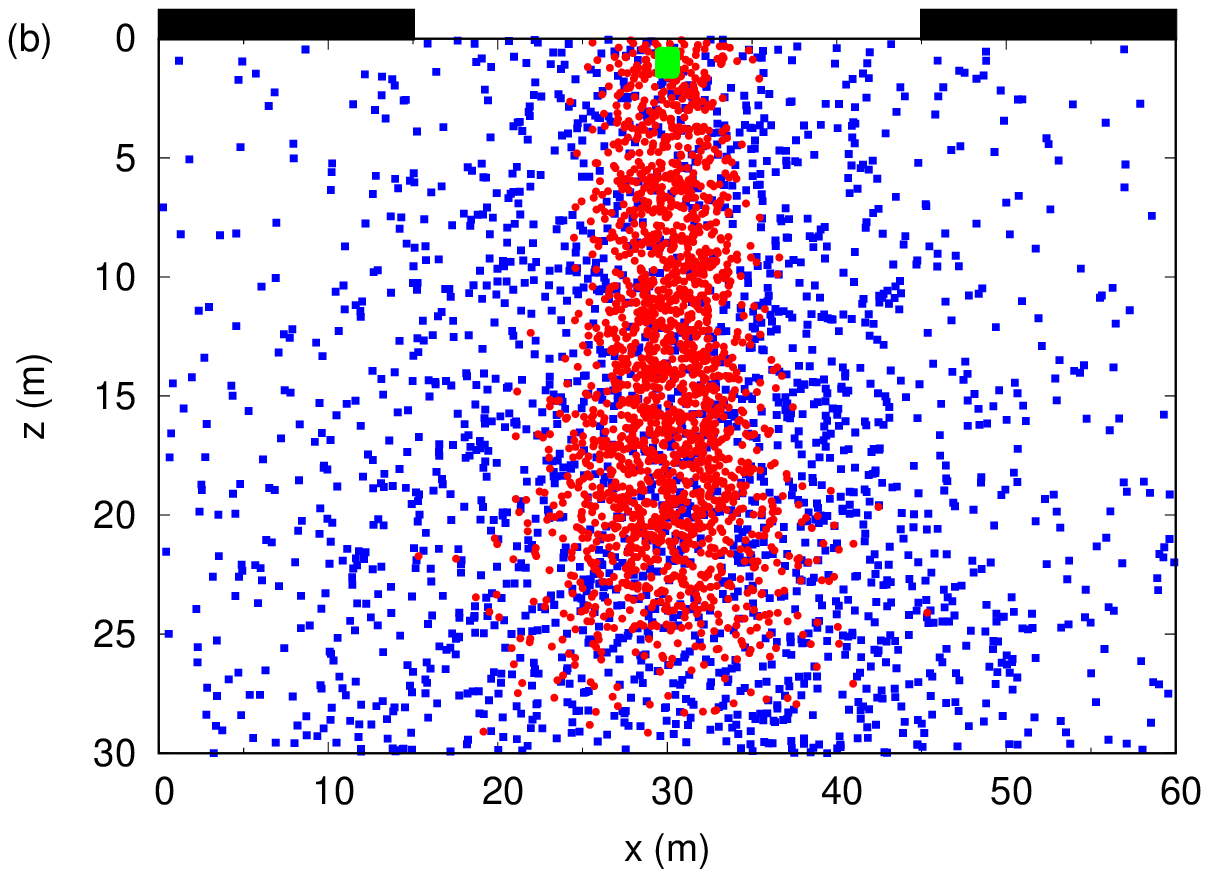}
\caption{Distribution of $N=2000$ particles at a time $t>0$ for $D=1$~cm$^2$~s$^{-1}$ (red circles) and $D=10$~cm$^2$~s$^{-1}$ 
(blue squares) in the LBD (a) and LBU (b) cases. In each case, at time $t=0$ all particles occupied a small area close to the top 
of the region receiving light (green circles).}
\label{fig:5}
\end{figure}
From a Lagrangian point of view, in the presence of diffusion, the complete equation of motion describing fluid transport is
$\dot{\bm{x}}=\bm{u}(\bm{x}(t))+v_{sink} \hat{\bm{z}}+\sqrt{2D}\bm{\eta}$, where the white noise term $\sqrt{2D}\bm{\eta}$
represents diffusion.
Sample distributions, at an arbitrary time $t>0$, of $N=2000$ particles evolving according to this equation
are shown in Fig.~\ref{fig:5} for $D=1$~cm$^2$~s$^{-1}$ (red circular points) and $D=10$~cm$^2$~s$^{-1}$ (blue square points). 
The two panels correspond to the LBD (Fig.~\ref{fig:5}a) and the LBU (Fig.~\ref{fig:5}b) case. 
In each of them, the particles were initially located in a small area close to the top of the illuminated region (green points). 
As expected, for larger $D$ particles spread more into the domain.
It is here worth emphasizing the lateral spreading: particles originally in a well-lit region are much more likely 
to reach a dark region, where reproduction is not possible, if their diffusivity is larger. 
In the LBD case this implies that, when $D$ is large enough, lateral diffusion can overcome the action of the sinking velocity, 
which helped maintaining particles in the upwelling reproductive region. In the LBU case, particles are also more easily brought 
to deadly regions by increased diffusivity, but this occurs while sinking contributes moving them downwards, 
where growth is also importantly reduced due to light absorption along the vertical. 
Therefore, the overall effect is weaker than in the previous case. 
This explains the asymmetry observed in the behavior of the per-capita growth rate as a function of the sinking speed,
with respect to the position of the obstacle blocking light. Indeed, in the LBD case, the positive action of $v_{sink}$ for growth 
is only barely visible for $D=10$~cm$^2$~s$^{-1}$ in Fig.~\ref{fig:4}b, as it is masked by the impact of diffusivity. 
On the contrary, in the LBU case, such a large value of the diffusion coefficient does not change the general phenomenology
(with respect to the $D=1$~cm$^2$~s$^{-1}$ case of Fig.~\ref{fig:4}a), as this was already unfavorable for growth, 
due to downward advection by both the sinking and vertical fluid velocities in the illuminated downwelling region.

We conclude this discussion by noting that the same reasoning can be made also considering transport backwards in time, 
which allows interpreting the results in terms of the generalization of Eq.~(\ref{eq:arL_sol1}) 
accounting also for diffusive transport:
\begin{equation}
\theta(\bm{x},t)=\langle \theta(\bm{x}(0),0)\ e^{\int_0^t g(\bm{x}(t')) dt'} \rangle_\eta \, .
\label{eq:ardL_sol}
\end{equation}
The above equation (see~\cite{abel2001front} for more details) states that the population density at time $t>0$ 
and at position $\bm{x}$ is constructed by advancing in time the reactive dynamics, starting from the value 
of the field $\theta(\bm{x}(0),0)$ at the Lagrangian origin $\bm{x}(0)$ of a trajectory ending at $\bm{x}$ at time $t$, 
and averaging over noise ($\eta$) realizations. Clearly, when the diffusion coefficient increases many more Lagrangian paths 
reaching $\bm{x}$ at time $t$ originate from regions where light cannot penetrate, bringing already dead material 
at the considered position at a later time.
In the LBD case, the effect is considerable, as the spatially and temporally varying per-capita growth rate $g(\bm{x}(t))$ 
in Eq.~(\ref{eq:ardL_sol}), at a location $\bm{x}$ in the illuminated (and upwelling) region, is now determined 
by the many trajectories coming from dark regions, which was not the case for smaller $D$.
In the LBU case, instead, also with small values of $D$ the contributions to $g(\bm{x}(t))$, with $\bm{x}$ in the illuminated 
(and downwelling) region, are essentially associated with trajectories starting in regions where light is blocked and the population 
has already died. Therefore, increasing $D$ does not alter the qualitative behavior of the system in this case.

\section{Conclusions}
\label{sec:concl}

We numerically investigated the dynamics of sinking phytoplankton in a stirred fluid layer in which light availability 
is spatially inhomogeneous, along both the vertical and the horizontal. Our system is assumed to be illuminated from above, 
as is typical in aquatic ecosystems receiving energy from the incoming solar radiation.  
Vertically, light intensity decreases due to absorption 
by water and phytoplankton self-shading. Horizontally, its penetration from the surface into the interior of the medium 
is critically affected by the presence of opaque obstacles 
at the top boundary.

We focused on the impact of advection, and more generally of the different transport processes occurring in the fluid, 
on the average biomass. In particular, we considered a 2D kinematic velocity field that provides an idealized representation 
of the flow rolls observed in thermal convection~\cite{solomon1988chaotic}, 
or in Langmuir circulation~\cite{craik1976rational,bees1997planktonic,thorpe2004langmuir} in the ocean mixed layer. 
Such flow pattern is characterized by both upwelling and downwelling regions, which play a key role on light-limited 
phytoplankton growth dynamics~\cite{taylor2011shutdown,lindemann2017dynamics,tergolina2021effects,lowry2018under}. 
Indeed, a general feature emerging from previous studies (taking into account the vertical decrease of light intensity 
but not its horizontal variability) is that vertical mixing due to convective motions reduces 
the growth of phytoplankton, and may even lead to extinction events~\cite{lindemann2017dynamics,tergolina2021effects}. 
This is largely due to the accumulation of the population in the downwelling regions associated with persistent 
large-scale vortical structures. Consequently, phytoplankton is efficiently transported to depth, where reproduction is inhibited 
by the lack of light~\cite{lindemann2017dynamics,tergolina2021effects}. Irregular fluid motions on smaller time and length scales 
only weakly affect this phenomenology~\cite{tergolina2021effects}.

Our goal in this work was to explore if and how the response of the system, in terms of average biomass and survival 
conditions, changes when the incident light is not uniform at the surface, a situation that is often encountered in natural 
environments. A remarkable instance of such a configuration is found in oceanic waters that are partially covered by ice. 
In spite of their relevance for primary-productivity estimates~\cite{arrigo2012massive,assmy2017leads}, under-ice blooms 
are still not well understood. 
Notwithstanding its idealized character, the model considered here can then be informative about the basic mechanisms controlling 
the interplay between transport and biological growth processes in polar, light-limited environments, in the prespective of properly 
representing fluid motions in more complex numerical models. 

Aiming to explore the simplest possible settings, we considered obstacles that fully absorb light and entirely cover either the 
downwelling region (LBD case) or the upwelling one (LBU case). We found that the main role of advective transport is to hinder 
phytoplankton growth. In both the LBD and LBU cases, the per-capita growth rate decreases with increasing flow intensity, and 
for sufficiently large advecting velocity, it can be become negative, meaning that the population goes extinct. 
While this behavior is qualitatively the same as the one observed in the absence of horizontal light-intensity modulation, 
now the quantitative effect depends on the position of the obstacles with respect to the regions of ascending and descending 
fluid motion. Phytoplankton survival is more difficult when the transmission of light is blocked over the upwelling 
flow region (LBU). 

The flow field alone cannot be responsible for the asymmetric response of the system in the LBD and LBU cases, 
as it can be understood by examining the advection-reaction dynamics in a Lagrangian reference frame.
We then explored the role of the sinking speed (due to the mass-density difference between phytoplankton organisms and water). 
Our analysis shows that, while typically much smaller than the flow intensity, the sinking speed is at the origin of the 
differences observed in the per-capita growth rates. We then provided a qualitative explanation of this phenomenon, which 
essentially depends on how the sinking speed combines with the vertical advecting velocity. 
In downwelling regions they sum, while in upwelling regions 
the sinking speed is subtracted from the flow. As a consequence, the residence time in the well-lit upper layer 
is increased in the upwelling region, relevant for the LBD case, and decreased in the downwelling region, 
relevant for the LBU case.   

Finally, we have shown that the combination of velocities in the vertical discussed above is delicate, 
and can be strongly affected by lateral diffusion. Indeed, in the present system, vertical and horizontal motions 
cannot be easily decoupled. Diffusion spreads the population on the horizontal, bringing it from regions that are favorable 
for reproduction to deadly ones. This effect is more marked in the LBD case, where the population is more concentrated 
at the surface (in the upwelling region, where sinking partially counteracts the flow). 
In the LBU case, instead, the dynamics are less sensitive to the intensity of diffusion, because in the (downwelling) 
illuminated region the population is already transported (by both the flow and sinking) to deeper waters 
where reproduction is also strongly reduced. 

To conclude, this study provided evidence that the onset of convection may be harmful to under-ice blooms, confirming 
previous findings~\cite{lowry2018under}. The most notable result, however, concerns the role of the phytoplankton sinking speed. 
While small, the latter has a measurable impact on the growth dynamics of the population and can even be critical for its survival, 
in partially ice covered waters undergoing convective motions. 
This fact may have ecological relevance, as different phytoplankton species have different densities and, hence, 
different settling velocity. 
The system dynamics revealed highly complex, also in such an idealized setting, due to the very subtle interplay 
bewteen the different transport mechanisms (fluid advection, sinking, diffusion) and biological growth, 
which limits the possibility to obtain an analytical understanding. An interesting perspective would be 
to carry out numerical simulations in a more realistic configuration for the fluid, by considering the dynamics 
of turbulent convection in the presence of a melting interface with the ice obstacle~\cite{esfahani2018basal}.

\section*{Acknowledgements}
\noindent We are grateful to M. Cencini for interesting discussions about the role of diffusion.

\section*{Data Availability Statement}
\noindent The data that support the findings of this study are available from the corresponding author upon reasonable request.


%
%


\end{document}